\documentclass[twocolumn,prl]{revtex4}

\ifx\pdfoutput\undefined
  \usepackage[dvips]{graphicx}
\else
  \usepackage[pdftex]{graphicx}
\fi

\usepackage{dcolumn}
\usepackage{amsmath}
\usepackage{latexsym}
\usepackage{units}

\begin{document}

\title[Short Title]{Self-shunted Al/AlO$_{\text x}$/Al Josephson
junctions}
\author{S.~V.~Lotkhov}
\email[Electronic mail:~]{Sergey.Lotkhov@ptb.de}
\author{E.~M.~Tolkacheva}
\author{D.~V.~Balashov}
\author{M.~I.~Khabipov}
\author{F.-I.~Buchholz}
\author{A.~B.~Zorin}
\affiliation{Physikalisch-Technische Bundesanstalt, Bundesallee
100, 38116 Braunschweig, Germany}%

\date{\today}

\begin{abstract}
Self-shunted aluminum Josephson junctions with high-transparency
barriers were fabricated using the shadow-evaporation technique
and measured at low temperatures, $T \approx \unit[25]{mK}$. Due
to high junction transparency, the $IV$-characteristics were
found to be of only small hysteresis with retrapping-to-switching
current ratio of up to 80~$\%$. The observed critical currents
were close to the Ambegaokar-Baratoff values (up to $\unit[80-100]
\%$). Good barrier quality was confirmed by the low subgap leakage
currents in the quasiparticle branches, which makes the
self-shunted Al junctions promising for application in integrated
RSFQ-qubit circuitry.

\end{abstract}

\maketitle

Operation of superconducting electronic devices often implies an
overdamped regime of Josephson dynamics \cite{Likh}. An important
example is given by a family of Rapid Single Flux Quantum (RSFQ)
devices (see, e.g., a review Ref.~\cite{RSFQ}) traditionally
based on resistively shunted junctions. A different type of
shunting is however preferred, when using RSFQ-circuitry as
control electronics for Josephson qubits \cite{RSFQ-Qubit}. For
keeping low the overall decoherence, low-noise behaviour of the
control RSFQ-circuit is required at qubit frequencies of the
order of ten GHz (see, e.g., the analysis of decoherence of the
flux qubit \cite{Chiao}). Due to large Niquist-Johnson
fluctuations at low frequencies, this requirement can hardly be
met with a standard low-ohmic linear shunting. Recently, several
realizations of frequency-dependent damping for Josephson
junctions have been proposed
\cite{Zorin-SIN,Zorin-microscopic,SIN2006,VTT}. In
Refs.~\cite{Zorin-SIN, Zorin-microscopic},
superconductor-insulator-normal metal (SIN) contacts were
suggested as nonlinear shunts for Josephson junctions and their
operation was analyzed in simple SFQ networks. A detailed study of
subgap conductivity in highly-transparent SIN-junctions was
presented in Ref.~\cite{SIN2006}. Another solution on the base of
RC-shunting was considered by Hassel {\it et al.} \cite{VTT}, to
be implemented in form of on-chip capacitors, added in series to
the standard ohmic shunts.

In this Letter, we suggest an alternative approach on the basis of
Al self-shunted Josephson junctions, making use of their highly
nonlinear quasiparticle branches. The power spectrum $P_{\text
{Iqp}}(\omega)$ of current fluctuations of such a junction in the
superconducting (S) state, $\left\langle V \right\rangle = 0$,
\begin{equation}
\label{Noise} P_{\rm Iqp} (\omega ) = (e/\pi )I_{\rm qp} (\hbar
\omega /e)\coth (\hbar \omega /2k_{\rm B} T),
\end{equation}
depends on the shape of its quasiparticle branch $I_{\rm qp}(V)$
\cite{Dahm} with the current suppressed at low temperatures,
$k_{\text B}T \ll \Delta$, in the subgap voltage range, $V
\lesssim V_{\text g}\equiv 2\Delta/e$. For Al junctions with the
superconducting energy gap $\Delta \sim \unit[200]{\mu eV}$, the
gap frequency, $\omega_{\text g} = 2\Delta / \hbar \sim 2\pi
\times \unit[10^{11}]{s^{-1}}$, far exceeds the typical qubit
frequencies, thus enabling low decoherence on a time scale of
qubit operation.

The fabrication routines for small Al/AlO$_{\text x}$/Al tunnel
junctions are well developed, basing on the shadow evaporation
technique \cite{Dolan}. In the early experiments, this technique
has also been implemented for submicron self-shunted tunnel
junctions composed of Pb-In/Pb \cite{Hu}. Moreover, in the last
years, it is widely applied for fabrication of Al-based qubits
(see, e.g., the review on Josephson qubits in Ref.~\cite{makhlin}
and references therein). The latter fact enables, in principle,
full on-chip integration of Al-RSFQ and Al-qubit circuitry, both
operating in sub-Kelvin temperature range.

The intrinsic damping properties of an autonomous Josephson tunnel
junction crucially depend on the barrier transparency, i.e., on
the specific barrier resistance $\rho = R_{\text N}A$, where
$R_{\text N}$ is the normal junction resistance and $A$ is the
junction area. Using for the critical current its
Ambegaokar-Baratoff (AB) value at $T = 0$, $I_{\text c} \approx
I_{\text c}^{\text {AB}} = \pi \Delta / 2eR_{\text N}$, one can
express the Stewart-McCumber damping parameter, $\beta_{\text c}
= 2e I_{\text c} R_{\text N}^2 C/\hbar = \pi \Delta\rho c /\hbar$
as a function of $\rho$, $\Delta$, and the specific junction
capacitance $c = C/A$. The overdamping condition, $\beta_{\text
c} \alt 1$, thus imposes an upper boundary on the barrier
transparency, $\rho \alt \hbar /(\pi \Delta c)$.

In experiment, non-hysteretic $IV$-curves \cite{Hu,Patel} and
high-frequency operation of an RSFQ circuit \cite{750GHz} built on
the self-shunted Nb/AlO$_{\text x}$/Nb junctions with
ultrahigh-transparency barriers, $\rho \sim \unit[1]{\Omega \times
\mu m^2}$ (corresponding to the high critical current density
$J_{\text c} \sim \unit[1-2]{mA / \mu m^2}$), have been
convincingly demonstrated. However the barrier quality was found
to be not so high, due to noticeable density of pin-hole defects
\cite{Patel,Kleinsasser} and, as a consequence, the considerable
subgap leakage currents \cite{Patel,Kleinsasser,Miller,Naveh}.

An important advantage of Al-based over Nb-based junctions for the
specific RSFQ-Qubit applications turns out to be a $\unit[7-8]{}$
times smaller value of $\Delta \approx \unit[180]{\mu eV}$
(corresponds to the zero-temperature BCS value $\Delta_0 =
1.76k_{\text B}T_{\text c}$ for $T_{\text c} = \unit[1.2]{K}$),
which allows non-hysteretic behaviour of more opaque junctions
with, typically, $\rho \alt \unit[30]{\Omega \times \mu m^2}$.
This decreases the risk of pin-hole defects and dramatically
weakens the subharmonic gap structure of multiple Andreev
reflection \cite{Kleinsasser}. In the reported here experiments
with Al junctions, we found the values of $I_{\text c}$ to be
slightly below their AB values. We also show that the
nonlinearity parameter $\eta \equiv \left[ G(0)R_{\text
N}\right]^{-1} \sim \unit[10-30]{}$, where $G(0)$ is the
zero-bias conductance, substantially exceeds the values $\eta
\sim 1$ measured in high-transparency Nb-junctions
\cite{Patel,Kleinsasser,Naveh}.

\begin{figure}[t]
\begin{center}
\leavevmode
\includegraphics[width=3.2in]{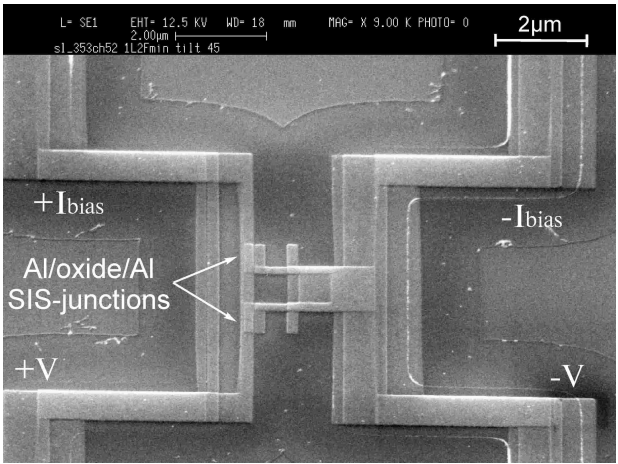}
\caption {SEM-micrograph of the two-junction SQUID. The junction
is formed between the first and the third layers of Al, whereas
the second layer (middle shadow of the mask) is not used in this
structure. The $IV$-curves were measured using a four-point
configuration.} \label{Fig1}
\end{center}
\end{figure}

In order to enable on-chip comparison between junctions with
different parameters, the sample was designed to include three
consecutive Al layers ("shadows"), 30, 40 and $\unit[20]{nm}$
thick. The layers were deposited {\em in situ\/} at three
different angles, $-22^{\circ}, +11^{\circ}$ and $+22^{\circ}$,
respectively, through the same PMMA/Ge/Copolymer mask,
$\unit[0.1/0.05/1.2]{\mu m}$-thick, with suspended bridges. The
first (the rightmost) shadow of Al was weakly oxidized at
$P_{\text O_2} =\unit[0.1]{Pa}$ for $\unit[5]{min}$ to form a
thin tunnel barrier. A set of junctions with two different
transparencies was formed in overlaps between the first and
either the second or, after prolonged oxidation of the first Al
shadow by adsorbed oxygen, the third layer (cf. similar design
for Al/Cu/Cu SIN-junctions in Ref.~\cite{SIN2006}). The results
are reported for two- and four-junction SQUIDs with a total
junction area $A \approx 2 \times \unit[(0.25 \times 1)]{\mu
m^2}$ and $A \approx 4 \times \unit[(0.25 \times 0.5)]{\mu m^2}$,
respectively; the SQUID loop area was $S \sim (1 \times
\unit[1)]{\mu m^2}$. A typical SQUID-device is shown in
Fig.~\ref{Fig1}. We found an on-wafer spread of tunnel resistances
of less than $\unit[10-20] \%$, indicating good junction
uniformity.

\begin{figure}[t]
\begin{center}
\leavevmode
\includegraphics[width=3.2in]{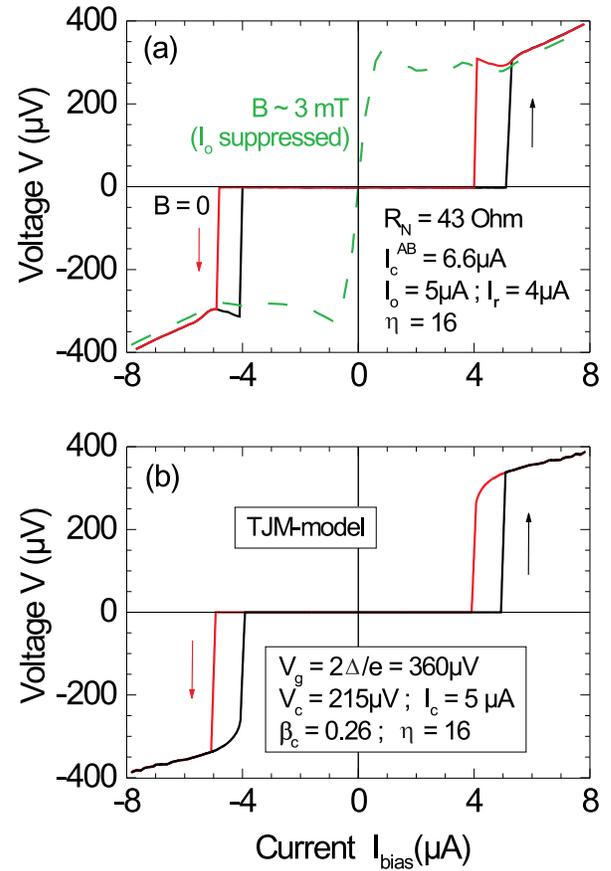}
\caption {(a) Typical current voltage characteristics of a
two-junction Al-SQUID, demonstrating a small remaining hysteresis
of the supercurrent. When the supercurrent is suppressed by
magnetic field, $BS = \Phi_{\text 0}(n+1/2)$, where $\Phi_{\text
0} = \unit[2.07 \times 10^{-15}]{Wb}$ is the flux quantum,
nonlinearity of the quasiparticle branch with $\eta \approx 16$
is observed. (b) TJM-model simulation of an SIS-junction with the
value of $\beta_{\text c}$ giving the best agreement with
experiment. The model does not take into account the junction
overheating effects, which were observed in experiment in form of
a local back-bending of $IV$-curves.} \label{Fig2}
\end{center}
\end{figure}

The $IV$-characteristics (see an example in Fig.~\ref{Fig2}(a))
were measured at $T \approx \unit[25]{mK}$ in the current-bias
mode. The obtained values of the switching current, $I_{\text o}
= \unit[5.2]{\mu A} \pm 10~\% $ ($J_{\text c} \sim \unit[10]{\mu
A / \mu m^2}$) and the retrapping current, $I_{\text r}=
\unit[3-4.4]{\mu A}$, correspond to a small hysteresis of 20 to
40~$\%$ and to an appreciable pair-current suppression factor
\cite{Zorin-tun}, $\alpha \equiv I_{\text o} / I_{\text c}^{\text
AB} \sim $~0.8. To find $I_{\text c}^{\text {AB}} \approx
\unit[6.4]{\mu A}$ we used an asymptotic junction resistance,
$R_{\text N} = \unit[44]{\Omega} \pm 10~\%$ (corresponding to a
value of $\rho \approx \unit[22] {\Omega \times \mu m^2}$). Using
a typical values of the specific junction capacitance, $c \approx
\unit[50-75]{fF/\mu m^2}$, we evaluate the McCumber parameter,
$\beta_{\text c} \sim \unit[0.8-1.2]{}$. In the junctions with
higher transparency (with $\rho = \unit[10-12]{\Omega \times \mu
m^2}$), formed between the first and the second shadows of Al, we
found $I_{\text o} = \unit[10-13]{\mu A}$, $I_{\text r}=
\unit[7-9]{\mu A}$, and $I_{\text c}^{\text {AB}}\approx
\unit[13]{\mu A}$. Despite the smaller value of $\beta_{\text c}
\sim \unit[0.5]{}$, a width of hysteresis was comparable to that
of the lower-transparency junctions.

A width of hysteresis of $\sim$~20~$\%$ found in most of the
junctions agrees roughly with the RSJN model using the values of
$\beta_{\text c} \sim \unit[0.5-1]{}$ and an effective
quasiparticle resistance at $V \alt V_{\text c}$, $R_{\text J}
\sim (\unit[2-4]) R_{\text N}$ (see, e.g., Fig.~4 in
Ref.~\cite{Prober}). However, more accurate calculations in the
frame of the tunnel junction microscopic (TJM) model (see, e.g.,
Ref.~\cite{Likh} and the references therein), accounting for the
experimental value of $\alpha$, produced the $IV$-curves, see
Fig.~\ref{Fig2}(b), in agreement with experiment, but
corresponding to a lower value of $\beta_{\rm c} \approx 0.3$.
The lower effective value of $\beta_{\rm c}$ indicates a stronger
damping in our experiment than the predicted intrinsic damping
due to quasiparticle tunnelling only. An additional damping
mechanism possibly arises due to a high-frequency shunting effect
of the junction by the on-chip wiring impedance. This effect
should be appreciable due to the relatively high values of
$R_{\text N}$ which are comparable to a typical microstrip-line
impedance, $R_{\text e} \sim \unit[10-100]{\Omega}$ (see, e.g.,
\cite{Holst}), and can explain a similar hysteresis in the
junctions of different transparency. An opposite effect of wiring
was a larger hysteresis, up to 44~$\%$, in those junctions which
had the shortest connecting paths, $\sim \unit[100]{\mu m}$ long,
to the wide leads of large capacitance. In practice, this
contribution can be considerable in the systems of submicron
tunnel junctions, but it can be reduced using, for instance,
adjacent biasing resistors.

The measurably larger retrapping current values can also be an
effect of thermal fluctuations, activated predominately in the
resistive (R) state due to the junction self-heating. Using the
model for noise-induced R$\to$S transitions,
Ref.~\cite{Ben-Jacob}, and the parameters of our junctions, we
found a realistic estimate for the effective temperature in the
R-state, $T^* \sim \unit[1]{K} < T_{\text c}$. This estimate is
consistent with the pronounced back-bending which is seen in the
$IV$-curves in Fig.~\ref{Fig2}(a) (cf., e.g.,
Ref.~\cite{Overheating}), and the observed $\sim 20 \%$ reduction
in the effective value of $V_{\text g}$ at $I \agt I_{\text o}$.

The quasiparticle branches of the $IV$-curves, like that shown by
the dashed line in Fig.~\ref{Fig2}, were measured in the SQUIDs
with a supercurrent suppressed by a magnetic field. The subgap
conductance was found to be low, $\left[ G(0) \right]^{-1} \sim
\unit[1]{k\Omega}$, which would allow low decoherence rates
\cite{Chiao} in all-Al RSFQ-qubit circuits. High values were
obtained for the nonlinearity parameter $\eta$, which were larger
than both in the Nb-based junctions \cite{Patel,Naveh} and in the
Al SIN-junctions \cite{SIN2006}. Similar to the case of
SIN-junctions, the values of $\eta$ were found to depend on the
junction configuration and size. In particular, the largest value
of $\eta \approx 32$, obtained in the four-junction SQUID,
exceeded $\eta \approx 16$, measured in the two-junction SQUID
with the two times larger junctions. In these two devices (not
shown), the junctions were simply formed as overlaps of the
straight perpendicular microstrips. A lower value of $\eta = 8.8$
was, however, found in a structure shown in Fig.~\ref{Fig1} with
one of electrodes made of "$\Gamma$"-shape (both right electrodes
of the junctions; we call this configuration an "electron
confinement shape"). The observed dependencies can be interpreted
following the argument developed in Ref.~\cite{HekNaz} for
describing an enhancement of the Andreev reflection processes due
to coherent two-electron (quasiparticle) diffusion in the junction
electrodes. (See for a more detailed discussion
Ref.~\cite{SIN2006}.)

To conclude, we investigated the damping in self-shunted
Josephson junctions of type Al/AlO$_{\text x}$/Al of high
transparency, fabricated using a shadow-evaporation technique.
The $IV$-characteristics were featured by a critical supercurrent
of the order of 10~$\mu A$ and a small hysteresis down to
20~$\%$. Strong subgap nonlinearity of the quasiparticle branches
was observed, which enables low decoherence in RSFQ-qubit
integrated circuitry on the basis of Al junctions.

The work was partially supported by the EU through projects
RSFQubit and EuroSQIP.

\end{document}